\documentclass[12pt]{article}
\usepackage{axodraw,bbold}

\catcode`@=12
\begin{document}
\thispagestyle{empty}
\begin{flushright} 
UCRHEP-T394\\ 
August 2005\
\end{flushright}
\vspace{0.5in}
\begin{center}
{\LARGE	\bf Efficacious Additions to the Standard Model\\}
\vspace{1.5in}
{\bf Ernest Ma\\}
\vspace{0.2in}
{\sl Physics Department, University of California, Riverside, 
California 92521, USA,\\
and Institute for Particle Physics Phenomenology, Department of Physics,\\ 
University of Durham, Durham, DH1 3LE, UK\\}
\vspace{1.5in}
\end{center}

\begin{abstract}\
If split supersymmetry can be advocated as a means to have gauge-coupling 
unification as well as dark matter, another plausible scenario is to 
enlarge judiciously the particle content of the Standard Model to achieve 
the same goals without supersymmetry.  A simple efficacious example is 
presented.
\end{abstract}

\newpage
\baselineskip 24pt

It has been proposed in the last year that supersymmetry could be split 
\cite{ss} so that scalar quarks and leptons are very heavy whereas 
gauginos and higgsinos remain light at the electroweak scale. This 
scenario ignores the hierarchy problem but has the virtue of retaining 
gauge-coupling unification as well as dark matter. It has inspired a 
large number of phenomenological studies.  On the other hand, if the 
goal is simply to preserve gauge-coupling unification and the appearance 
of dark matter, then a judicious choice of particle content beyond that 
of the Standard Model (SM) works just as well.  Such an alternative may 
have very different predictions at the TeV scale and should not be 
overlooked in this context.  A simple efficacious example is presented 
below.

It has been known for a long time that the particle content of the SM 
does not lead to unification of the three gauge couplings of the standard 
$SU(3)_C \times SU(2)_L \times U(1)_Y$ gauge group, whereas that of the 
Minimal Supersymmetric Standard Model (MSSM) does \cite{bs04}. To understand 
how this works, consider the one-loop renormalization-group equations 
governing the evolution of these couplings with mass scale:
\begin{equation}
{1 \over \alpha_i(M_1)} - {1 \over \alpha_i(M_2)} = {b_i \over 2 \pi} 
\ln {M_2 \over M_1},
\end{equation}
where $\alpha_i = g_i^2/4\pi$ and the numbers $b_i$ are determined by 
the particle content of the model between $M_1$ and $M_2$.  In the SM 
with one Higgs scalar doublet, these are given by
\begin{eqnarray}
SU(3)_C &:& b_C = -11 + (4/3) N_f = -7, \\
SU(2)_L &:& b_L = -22/3 + (4/3) N_f + 1/6 = -19/6, \\
U(1)_Y &:& b_Y = (4/3) N_f + 1/10 = 41/10,
\end{eqnarray}
where $N_f = 3$ is the number of quark and lepton families and $b_Y$ 
has been normalized by the well-known factor of 3/5.  In the MSSM with 
two Higgs superfields, the shifts in $b_i$ from those of the SM are given by
\begin{eqnarray}
\Delta b_C &=& (2/3) N_f + 2, \\
\Delta b_L &=& (2/3) N_f + 13/6, \\
\Delta b_Y &=& (2/3) N_f + 1/2.
\end{eqnarray}
Since the scalar quarks and leptons are contained in the terms proportional 
to $N_f$, they do not change the relative values of $b_i$. Hence they do 
not affect the unification condition
\begin{equation}
\alpha_C(M_U) = \alpha_L(M_U) = (5/3)\alpha_Y(M_U) = \alpha_U,
\end{equation}
or the value of $M_U$; they only change the value of $\alpha_U$.
Split supersymmetry is essentially the scenario where
\begin{equation}
\Delta b_C = 2, ~~~ \Delta b_L = 13/6, ~~~ \Delta b_Y = 1/2,
\end{equation}
which is realized by adding to the SM new fermions transforming as 
(8,1,0), (1,3,0), (1,1,0), (1,2,$\pm 1/2$), and a second scalar Higgs 
doublet.  Suppose instead we add two complex scalar octets $\zeta, \zeta' 
\sim (8,1,0)$, one Majorana fermion triplet $f \sim (1,3,0)$ as well as 
one complex scalar triplet $s \sim (1,3,0)$, then
\begin{equation}
\Delta b_C = 2, ~~~ \Delta b_L = 2, ~~~ \Delta b_Y = 0,
\end{equation}
which mimics Eq.~(9) to a large extent. Assuming that these new particles 
appear at around $M_X$, then the unification condition of Eq.~(8) implies
\begin{equation}
{1 \over \alpha_C(M_Z)} = {3 \over 158} \left[ {91 \over \alpha_L(M_Z)} - 
{23 \over \alpha_Y(M_Z)} \right] + {15 \over 158 \pi} \ln {M_X \over M_Z},
\end{equation}
\begin{equation}
\ln {M_U \over M_Z} = {30 \pi \over 79} \left[ {3 \over 5 \alpha_Y(M_Z)} - 
{1 \over \alpha_L(M_Z)} \right] - {30 \over 79} \ln {M_X \over M_Z}.
\end{equation}
Using the input \cite{data}
\begin{eqnarray}
\alpha_L(M_Z) &=& (\sqrt 2/\pi) G_F M_W^2 = 0.0340, \\
\alpha_Y(M_Z) &=& \alpha_L(M_Z) \tan^2 \theta_W = 0.0102,  
\end{eqnarray}
and
\begin{equation}
0.115 < \alpha_C(M_Z) < 0.119,
\end{equation}
we find
\begin{equation}
1800~{\rm GeV} > M_X > 500~{\rm GeV},
\end{equation}
and
\begin{equation}
5.1 \times 10^{16}~{\rm GeV} < M_U < 8.3 \times 10^{16}~{\rm GeV}.
\end{equation}
These are certainly acceptable values for new particles at the TeV scale 
and the proper suppression of proton decay.

The new particles assumed are adjoint representations of $SU(3)_C$ or 
$SU(2)_L$.  They could for example be components of the adjoint 
\underline{24} representation of $SU(5)$.  There is no fundamental 
understanding of why they are light, but the other components are heavy, 
just as there is no fundamental understanding of why the scalar $SU(2)_L$ 
doublet components of the \underline{5} representation of $SU(5)$ are light, 
but the $SU(3)_C$ triplet components are heavy in the canonical $SU(5)$ 
model of grand unification.

We now come to the phenomenology of the new particles.  The $SU(3)_C$ scalar 
octets $\zeta, \zeta'$ may be thought of as scalar gluons.  They decay into 
two vector gluons in one loop, i.e. $\zeta \to \zeta \zeta \to g g$, etc.  If 
kinematically allowed, they would be 
produced copiously at the Large Hadron Collider (LHC) and easily detected.  
As far as $\Delta b_C = 2$ in Eq.~(10) is concerned, $\eta$ and $\zeta$ may be 
replaced by a single Majorana fermion octet, i.e. the gluino in the MSSM, 
but then the latter would be absolutely stable without scalar quarks, which 
is not acceptable cosmologically. [In split supersymmetry, the scalar quarks 
are still present, only much heavier.]

Instead of the required two Higgs superfields in the MSSM, this model 
has only the single Higgs boson of the SM.  Hence its mass is not 
constrained by the MSSM and may exceed the latter's current upper bound of 
about 127 GeV \cite{127}.

The Majorana fermion triplet $(f^+,f^0,f^-)$ are like the three $SU(2)_L$ 
gauginos of the MSSM.  However, there is no $U(1)_Y$ gaugino here.  
After all, it contributes nothing to $b_C$, $b_L$, or $b_Y$, so it is 
not necessary for gauge-coupling unification.  In the MSSM, the gauginos 
couple to the Higgs scalars through the higgsinos, but here there are 
no higgsinos. However, since the left-handed leptons $(\nu,l)$ are 
doublets, $f$ can couple to them through the SM Higgs doublet, which 
would allow $f$ to decay and disqualify it from being considered as dark 
matter.  The solution here is the same as in the MSSM.  We simply assign 
a conserved quantity to $f$, say a multiplicative parity under which $f$ 
is odd and all other particles are even, in exact analogy to $R$-parity 
in the MSSM.

The complex scalar triplet $(s^+,s^0,s^-)$ couples to the SM Higgs doublet 
$(\phi^+,\phi^0)$ according to
$$ s^+ \phi^- \phi^0 + s^0 (\bar \phi^0 \phi^0 - \phi^- \phi^+)/\sqrt 2 - 
s^- \bar \phi^0 \phi^+,$$
which means that $s^0$ will acquire a nonzero vacuum expectation value. 
This has to be smaller than about a GeV to satisfy the precision 
electroweak measurements of the $W$ and $Z$ masses. It is nevertheless 
important for the phenomenology of dark matter because it allows $f^\pm$ 
to be heavier than $f^0$ from the coupling 
$$s^+ f^0 f^- + s^0 f^- f^+ + s^- f^+ f^0,$$
so that $f^\pm$ may decay weakly into $f^0$ plus a virtual $W^\pm$ 
which becomes a quark-antiquark pair or lepton-antilepton pair. 
Thus $f^0$ (the analog of the neutral wino in the MSSM) is the 
candidate for dark matter in this model.

The idea that particles beyond those of the SM can restore gauge-coupling 
unification and suppress proton decay is not new \cite{bm84}, but with the 
possible abandonment of low-energy supersymmetry as the only acceptable 
scenario below the TeV scale, new alternatives should be explored. 
One such simple example has been presented in this paper, with 
experimentally verifiable predictions very different from those of 
the MSSM or split supersymmetry.

This work was supported in part by the U.~S.~Department of Energy under 
Grant No. DE-FG03-94ER40837.

\bibliographystyle{unsrt}

\end{document}